# Fabry-Pérot fiber cavity refractive index sensing via linewidth tracking in the broken PT-symmetric region

FAIZA IFTIKHAR,[1,2] RAJA AHMAD,[3,4] AND M. IMRAN CHEEMA[1,*]

[1]*Electrical Engineering Department, Syed Babar Ali School of Science and Engineering, Lahore University of Management Sciences (LUMS), Lahore 54792, Pakistan*
[2]*Current affiliation: Electrical Engineering Department, Lahore College for Women University, Lahore, Pakistan*
[3]*Optical Solutions Group, Molex LLC., 440 US-22E, Bridgewater, NJ, 08807, USA*
[4]*Current affiliation: Cisco Systems Inc., 3700 Cisco Way, San Jose, CA, 95134, USA*
*\*Corresponding author: imran.cheema@lums.edu.pk*

**Parity-time (PT) symmetric optical sensors operating around exceptional points have recently gained much attraction, offering an unparalleled high sensitivity in measuring small perturbations. In the past, most of the PT-symmetric sensors have been based on tracking the *mode-splitting* that arises due to a perturbation-induced change in coupling strength between two subcavities of the PT-symmetric system. We design a linear fiber Fabry-Pérot and coupled cavities sensor, tailored to operate in the broken PT-symmetric region. We explore a new sensing metric – that is, the mode's *linewidth change* as a function of perturbation-induced changes in the loss within one of the subcavities of the PT-symmetric system. The coupling strength between the two subcavities remains unchanged in our proposed sensor. Supported by a mathematical formulation, we find that the full-width-half-maximum (FWHM) of the cavity resonances exhibits a square root dependence on the refractive index (RI) change in one of the subcavities. The proposed fiber cavity refractive index sensor has a maximum sensitivity of $2.26 \times 10^7$ GHz/RIU and a lowest detection limit of $10^{-9}$ RIU, widely outperforming the comparable cavity sensors subject to the same refractive index change, gain, and loss settings.**

The fiber Fabry-Pérot etalons (FFPEs) have been used widely in various domains such as lasers [1], wavelength filters [2], and optomechanical devices [3]. Their simple structure, low cost, and immunity to electromagnetic radiation make FFPEs a desirable platform for sensing applications [4, 5]. For a widescale adoption across industries, FFPEs remain to be developed with improved sensitivity and a lower detection limit to meet the relevant performance requirements. In the past, in a single-cavity FFPE platform, resonant frequency and FWHM changes have been employed as sensing metrics [6]. Researchers have also used active cavities to enhance the sensitivity of FFPE sensors [7]. To improve the sensing sensitivity further, a plausible platform consists of coupled gain and loss cavities that are arranged in a non-Hermitian parity-time symmetry (PTS) configuration [8]. This enables the resulting FFPE to perform exceptional point sensing, which is not explored in detail for the coupled fiber cavities.

The exceptional point phenomenon has recently generated increased interest in optical sensing within small footprint configurations [9]. Researchers have designed various optical systems such as WGM micro-cavities [10], ring resonators [11], gyroscopes [12], and resonant optical tunneling effect resonators [13] for refractive index, temperature, and pressure sensing. Although in most proposed schemes, the sensing was performed by relating the amount of the mode-splitting to the change in the coupling strength between coupled resonators, there were, however, some exceptions. For example, in one of the gyroscope-based works [14], researchers analyzed the mode's FWHM changes as a function of perturbations-induced coupling strength changes in coupled ring resonators.

The exceptional point sensing in optical fiber-based cavities remains largely unexplored. In one of the fiber loop resonator works, researchers sensed the mode-splitting as a function of coupling strength changes induced by temperature changes [15]. The linear fiber cavities are more sensitive as the propagating mode samples perturbing regions twice than loop cavities [7]. However, in linear fiber cavities, perturbations-introduced coupling strength changes are challenging to exploit for sensing purposes.

In the present work, we provide the first proposal, to our knowledge, for using coupled and linear fiber cavities in the broken PT-symmetric region and introduce a new sensing metric of FWHM change as a function of induced loss in one of the cavities. Notably, the coupling strength remains constant in our designed sensor. The sensor consists of a tapered fiber in one of the cavities (loss-cavity) as a sensing head that responds to external refractive index perturbations. Consequently, changes are induced in the gain-loss balance within the PT-symmetric cavity without affecting the joint coupling strength of the system. We demonstrate that our FFPE sensor exhibits a square-root dependency of FWHM on the loss change in the cavity due to refractive index change. We also show that, compared to a single-cavity fiber sensor, the FWHM enhancement is significantly larger in the proposed PT-symmetric system. We determine that our sensor has a maximum sensitivity of $2.26 \times 10^7$ GHz/RIU and the lowest detection limit of $10^{-9}$ RIU, offering a promising platform for physical and chemical sensing in a wide range of applications.

A linear and coupled FFPE is a non-Hermitian system with a complex-valued spectral response [16]. In the complex spectral response, the imaginary part of the complex eigenvalue (resonant frequency) represents the decay constant related to finite spectral FWHM at the resonance frequency, whereas the real part represents the eigenenergy. Similarly, the eigenstates are the resonant modes of the optical non-Hermitian system. The fundamental work by Bender et al. [17] shows that non-Hermitian systems can also have real spectra if they follow a unique symmetry called PT-symmetry, i.e., [H, PT]=0 [18]. In the PT-symmetric FFPE, when all the eigenvalues and eigenstates of the system coalesce for a particular parameter space, the system is at an exceptional point. Eigenvalues in the broken symmetry region respond strongly to the external perturbation near an exceptional point.

We consider a fiber Fabry-Perot sensor having two coupled cavities with end reflectivities $R_1=R_3=0.8$ and equal optical path lengths $L_1 = L_2 = 5$ cm, as shown in Figure 1. We design one of the cavities as a gain medium by incorporating optical amplification and the other cavity as a loss medium using a tapered fiber of diameter 1 $\mu m$. Based on the coupled mode equations, the Hamiltonian of an FFPE sensor comprising two coupled cavities (one representing net gain $B_1$ and the other with net loss $B_2$) of equal length, with resonance frequencies $\omega_1$ and $\omega_2$, can be expressed as $H = \begin{pmatrix} -\omega_1 - i\frac{B_1}{2} & J \\ J & -\omega_2 - i\frac{B_2}{2} \end{pmatrix}$, where $J$ is the joint coupling strength between the two cavities [19]. The net gain $B_1$ is the sum of losses due to finite reflectivity $R_1$ of $FBG_1$ and gain due to amplification in the gain cavity. Similarly, the net loss $B_2$ is the cumulative loss due to $FBG_3$'s finite reflectivity $R_3$ and loss due to a tapered fiber segment in the loss cavity, as shown in Figure 1.

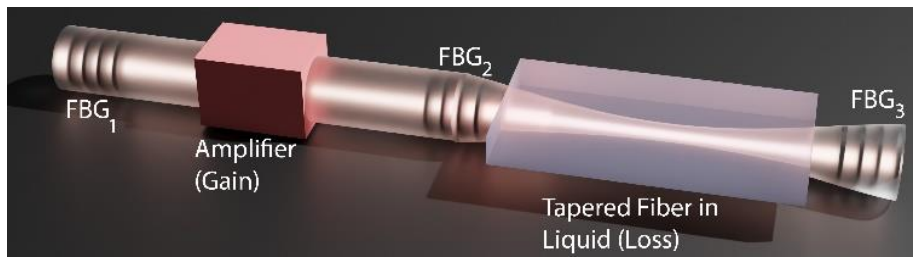

**Fig. 1.** Proposed FFPE sensor with coupled gain and loss cavities. The amplifier in the gain cavity compensates for all the system losses. The tapered fiber immersed in the liquid adds loss in the loss-cavity through the evanescent field leakage. The FFPE sensor has fixed reflectors, i.e., fiber Bragg gratings (FBGs) with reflectivities $R_1$, $R_2$, and $R_3$.

Using the finite element method (FEM), we determine the loss induced into the loss cavity by the tapered fiber when immersed in water. We then determine $B_2$= 14.742 GHz. To achieve an exceptional point in the FFPE sensor, we set $B_1$=14.742 GHz to balance gain and loss in the cavities to satisfy the PT-symmetry condition ($B_1 = -B_2$). We then select joint coupling strength, $J_{th} = \frac{1}{2}|B_2| = \frac{1}{2}|B_1| = 7.371$ GHz, to achieve the exceptional point where the system has the same resonance frequencies for both the modes, i.e., $\omega_1 = \omega_2 = \omega_o$. We determine the $R_2$ of the middle reflector using $J_{th} = 1.5 \times 10^8 \sqrt{\frac{1-R_2}{L_1 L_2}}$ [20]. At the exceptional point, the eigenfrequencies of the FFPE sensor with balanced gain and loss in the cavities can be expressed as:

$$\xi_{1,2} = \omega_0 \pm \frac{1}{4}\sqrt{16J_{th}^2 - (\omega_1 - \omega_2)^2} \qquad (1)$$

With a fixed joint coupling strength ($J_{th}$), when the loss in the cavity changes via tapered fiber, the system can experience three regimes: (i) PT-symmetry broken regime with $|B_1 - B_2| > J_{th}$ where the eigenfrequencies have equal real parts (no mode-splitting) but have equal and opposite imaginary parts leading to FWHM broadening, (ii) PT-symmetric unbroken regime with $|B_1 - B_2| < J_{th}$ where eigenfrequencies are real, leading to mode-splitting, and (iii) The exceptional point (EP) where eigenfrequencies and eigenvectors coalesce. These three regimes are shown in Figure 2.

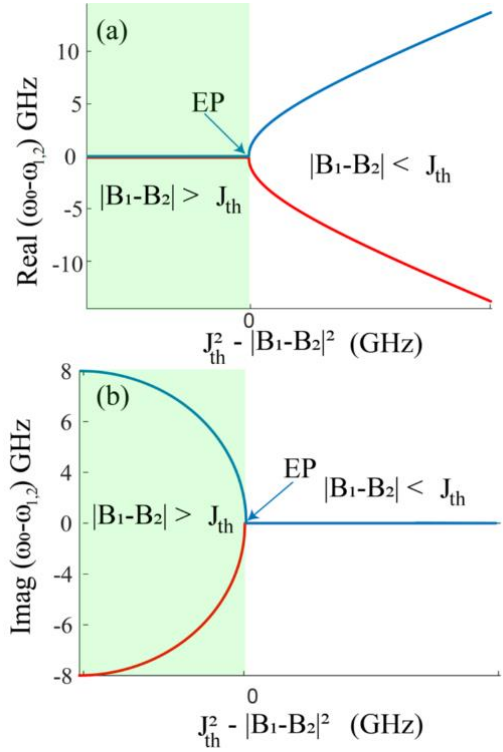

**Fig. 2.** The evolution of real and imaginary parts of eigenfrequencies of the two modes in an FFPE sensor when a loss in the loss cavity is varied. The exceptional point is located at $J_{th}$ = 7.371 GHz, having balanced gain and loss, i.e., B1 = −B2 = 14.7 GHz. (a) The real part of eigenfrequencies in the broken PT-symmetric region. (b) The imaginary part of eigenfrequencies in the broken PT-symmetric region.

After parking the coupled FFPE system at the exceptional point, we vary the refractive index of water by $10^{-9}$ to $10^{-4}$ RIU in which the tapered fiber is assumed to be immersed. Using FEM, we determine the evanescent field for the tapered fiber for each refractive index change and calculate the corresponding loss $B_2$ experienced by the FFPE system. The perturbations due to varying loss shift the system into the broken PT-symmetric region where gain and loss in the cavities are no more balanced. The resonance frequency of the loss cavity changes from $\omega_o$ to $\omega_o + \Delta\omega$ resulting in the following change in the eigenfrequencies of the system as determined from the Hamiltonian:

$$\Delta\omega_{PTS-b} = \sqrt{\frac{\Delta\omega \sqrt{\Delta\omega^2 + 16J_{th}^2} + \Delta\omega^2}{2}} + i\sqrt{\frac{\Delta\omega \sqrt{\Delta\omega^2 + 16J_{th}^2} - \Delta\omega^2}{2}} \quad (2)$$

The $\Delta\omega_{PTS-b}$ represents the eigenfrequencies difference in the broken PT-symmetric region, where $\Delta\omega = 4\pi c\, \Delta n l / K\lambda_o^2$ [21] and K is an integer whose value is 2237419 at 1550 nm.

The imaginary part of $\Delta\omega_{PTS-b}$ provides information about the FWHM of the system. In the broken PT-symmetric region, near the exceptional point, where $\Delta\omega^2 \ll 16J_{th}^2$, we can write the equation of FWHM ($\Delta\omega_{FWHM}$) in terms of the refractive index change ($\Delta n$) as

$$\Delta\omega_{FWHM} = \sqrt{\frac{\Delta 8\pi c \Delta n' l J_{th}}{K\lambda_o^2}} \quad (3)$$

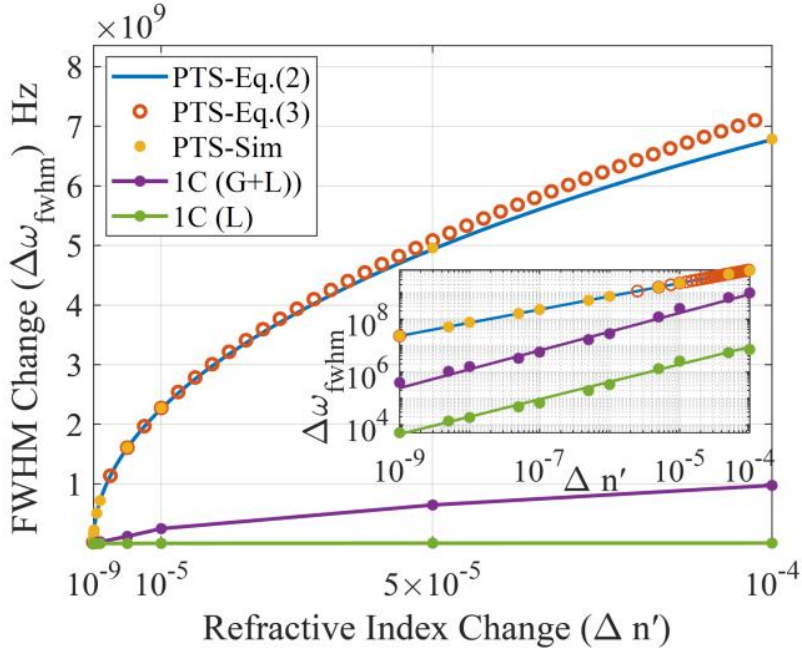

**Fig. 3.** FWHM change comparison of an FFPE sensor with coupled gain and loss cavities with a single-cavity (1C) sensor having balanced gain and loss and a single-cavity (1C) sensor with loss only. The inset figure is the log-log plot.

The numerical simulations in Figure 3 show FWHM enhancement in the broken PT-symmetric region of an FFPE sensor for $10^{-9}$ to $10^{-4}$ RIU range of refractive index change. The FWHM increases sharply with a square-root dependence on the refractive index change in the broken PT-symmetric region. The blue plot represents theoretically predicted FWHM ($\Delta\omega_{FWHM}$) using Eq. (2). Whereas the red circle plot shows our derived approximate FWHM using Eq. (3). The approximate FWHM overlaps the theoretically predicted FWHM near the exceptional point in the broken PT-symmetric region. However, when we move away from the exceptional point, the approximate FWHM no longer represents the system's FWHM.

To validate our derived FWHM expression, we evaluate the FWHM of the transmission profiles for selected values of the refractive index change. We determine the transmission profiles for the assumed values of refractive index change using the electromagnetic model of a multicavity Fabry-Pérot etalon presented in [21]. In Figure 3, the yellow-filled circles represent the FWHM of the transmission profiles determined through Lorentzian fit for selected values of refractive index change. In the near exceptional point region where the refractive index change is ultra-small of order $10^{-9}$ to $10^{-6}$ RIU, the yellow-filled circles are mapped entirely to the plot of approximate FWHM change. As we move away from the exceptional point, the numerically simulated FWHM deviates from the approximate FWHM using Eq. (3). This verifies the concept of FWHM change enhancement due to ultra-small refractive index change near the exceptional point in the broken PT-symmetric regime.

Now, we compare our proposed FFPE sensor operated in the broken PT-symmetric region with (i) a traditional single-cavity FFPE sensor with loss only and (ii) a single-cavity FFPE sensor with gain and loss. For a fair comparison, we make some assumptions, i.e., (i) the end reflectors' reflectivities will be the same for all the sensors, (ii) the net optical path lengths of the fiber cavities for all the sensors will be equal, and (iii) all the sensors will have identical tapered fiber segments to add absorption loss. The numerical simulations represented by green in Figure 3 show that the FWHM change is directly proportional to the refractive index change in a traditional single-cavity sensor without gain. However, when we add equal gain as the loss in the cavity due to the tapered fiber, the slope of FWHM change increases for the same refractive index change as in the single-cavity sensor without gain, as shown by the purple plot in Figure 3. With the square-root dependence of FWHM change on the refractive index change, our proposed FFPE sensor in the broken PT-symmetric region outperforms the traditional single-cavity FFPE sensor with a maximum sensitivity of $2.26 \times 10^7$ GHz/RIU and detection limit of $10^{-9}$ RIU. The inset is the log-log plot of FWHM curves of the three sensors compared in Figure 3. In the log-log plots, the single-cavity FFPE sensor in both the cases mentioned in (i) and (ii) shows a linear response. In contrast, the log-log plot of our proposed FFPE sensor in the broken PT-symmetric region shows a linear response with a half-slope as compared to (i) and (ii) due to square root dependence in the refractive index change on a linear scale.

The range of refractive index change for which our derived FHWM expression holds depends on the length of the tapered fiber, which is also the length of the loss cavity. We analyze the shift in refractive index change limit for different tapered fiber lengths in the 1 cm to 10 cm range, as shown in Figure 4. We notice that for the given lengths, the deviation in the refractive index change limit for which our expression of FWHM change holds is small. Hence, we can take any value of the tapered fiber length from 1cm to 10 cm.

In an FFPE sensor operated in the broken PT-symmetry region, the change in the real part of the eigenfrequencies becomes zero leading to no mode-splitting. In contrast, the FWHM of the sensor enhances due to the square root topology of the complex energy near the exceptional point. For an FFPE sensor, one can achieve a larger sensitivity enhancement for smaller perturbations near the exceptional point in a broken

PT-symmetric region. The FWHM change as a function of the refractive index is used as a sensing tool in a single-cavity fiber sensor. However, previously, it has not been used for sensitivity enhancement in an FFPE sensor with coupled cavities. This paper uses FWHM change as a function of the refractive index in the coupled-cavity fiber sensor designed to operate in the broken PT-symmetry region. The novelty of our work is that we have tuned the loss in the sensor cavity by independently changing the refractive index in the tapered fiber without changing the coupling strength of the gain-loss subcavities. The coupling strength depends on the reflectivity of the central FBG (Figure 1), which is fixed and independent of gain and loss tuning. In the coupled ring resonators, where mode-splitting enhancement is generally used as a sensing technique, a loss is added to the resonator while perturbing the coupling strength [22].

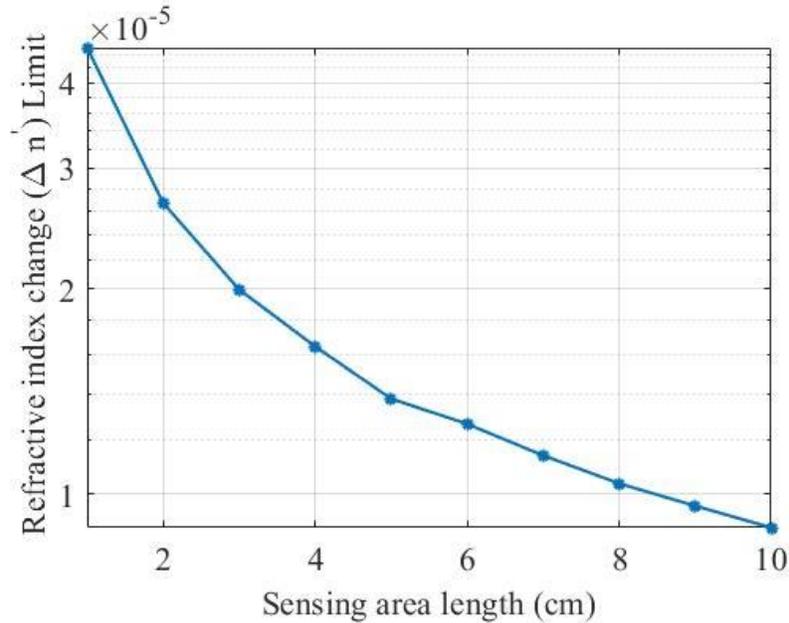

**Fig. 4.** An analysis of tapered fiber (sensing area) length and refractive index change (Δn) limit for which expression Eq. (3) holds.

An experimental implementation of the proposed refractive index sensor can be realizable by using off-the-shelf fiber components. The loss-cavity can be built by joining a tapered fiber with the FBGs on single-mode fibers (SMF28) with a central wavelength of 1550 nm. The gain-cavity can be designed using an amplifier [7] or a custom-made erbium-doped fiber amplifier (EDFA) with the specified amplification in dB required to balance loss in the loss-cavity. A tunable laser source and an optical detector can be used as input and output devices for the sensor.

The challenges we may face while implementing the setup are noise sources in the sensor, including the amplifier's nonlinearity, gain fluctuations, and ambient temperature effects on FBGs and the analyte. One can minimize the FBGs reflectivities' shift due to the temperature variations by experimenting in a temperature-controlled environment and choosing the FBGs with narrow bandwidth to ensure the wavelength scanning range is small as compared to the optical amplifier bandwidth. One can also implement the proposed sensor on a chip with an onboard amplifier and tapered fiber cavity-based sensing head.

In summary, we present an ultra-sensitive FFPE refractive index sensor operating in the broken PT-symmetry region. We find that a significant enhancement in FWHM can be induced by an ultra-small refractive index change in the sensor's fiber subcavity. We validate FWHM enhancement and its dependence on the discretely added loss in the fiber subcavity through theoretical analysis and numerical simulations while keeping the joint coupling strength fixed. We compare the FWHM change in our FFPE sensor, having coupled gain and loss subcavities arranged in a PT-symmetric configuration, with a traditional single-cavity FFPE sensor having a similar amount of gain and loss. Due to the square-root dependence of FWHM on the refractive index change in our FFPE sensor, the measurement sensitivity is increased manyfold with respect to that of the single-cavity FFPE sensors that exhibit a linear dependence of FWHM on RI change. Our results indicate that an FFPE sensor operating in the broken PT-symmetric region outperforms a conventional single-cavity fiber sensor with a maximum sensitivity enhancement of $2.264 \times 10^7$ GHz/RIU for the $10^{-9}$ to $10^{-4}$ RIU range of refractive index change. We anticipate that the present work will find various liquid and gas phase sensing applications involving fiber cavities.

**Funding.** Syed Babar Ali Research Award, 2020

**Disclosures.** The authors declare no conflicts of interest.

**Data availability.** Data underlying the results presented in this paper are not publicly available at this time but may be obtained from the authors upon reasonable request.

# References

[1] C. Liao, T. Hu, and D. Wang, "Optical fiber Fabry-Perot interferometer cavity fabricated by femtosecond laser micromachining and fusion splicing for refractive index sensing," *Optics express,* vol. 20, no. 20, pp. 22813-22818, 2012.

[2] J. Zhang, D. Hua, Y. Ding, and Y. Wang, "Digital synthesis of optical interleaver based on a solid multi-mirror Fabry–Perot interferometer," *Applied Optics,* vol. 56, no. 36, pp. 9976-9983, 2017.

[3] L. Belsito *et al.*, "Micro-Opto-Mechanical technology for the fabrication of highly miniaturized fiber-optic ultrasonic detectors," in *2011 16th International Solid-State Sensors, Actuators and Microsystems Conference*, 2011: IEEE, pp. 594-597.

[4] P. Zhang *et al.*, "Cascaded fiber-optic Fabry-Perot interferometers with Vernier effect for highly sensitive measurement of axial strain and magnetic field," *Optics express,* vol. 22, no. 16, pp. 19581-19588, 2014.

[5] U. Ullah, M. Yasin, A. Kiraz, and M. I. Cheema, "Digital sensor based on multicavity fiber interferometers," *JOSA B,* vol. 36, no. 9, pp. 2587-2592, 2019.

[6] Y. Liu and S. Qu, "Optical fiber Fabry–Perot interferometer cavity fabricated by femtosecond laser-induced water breakdown for refractive index sensing," *Applied optics,* vol. 53, no. 3, pp. 469-474, 2014.

[7] U. Ullah and M. I. Cheema, "Phase shift-cavity ring down spectroscopy in linear and active fiber cavities for sensing applications at 1550 nm," *IEEE Sensors Journal,* vol. 21, no. 12, pp. 13335-13341, 2021.

[8] M.-A. Miri and A. Alù, "Exceptional points in optics and photonics," *Science,* vol. 363, no. 6422, p. eaar7709, 2019.

[9] Q. Zhong, J. Ren, M. Khajavikhan, D. N. Christodoulides, Ş. Özdemir, and R. El-Ganainy, "Sensing with exceptional surfaces in order to combine sensitivity with robustness," *Physical review letters,* vol. 122, no. 15, p. 153902, 2019.

[10] W. Chen, Ş. Kaya Özdemir, G. Zhao, J. Wiersig, and L. Yang, "Exceptional points enhance sensing in an optical microcavity," *Nature,* vol. 548, no. 7666, pp. 192-196, 2017.

[11] S. Sunada, "Large Sagnac frequency splitting in a ring resonator operating at an exceptional point," *Physical Review A,* vol. 96, no. 3, p. 033842, 2017.

[12] M. P. Hokmabadi, A. Schumer, D. N. Christodoulides, and M. Khajavikhan, "Non-Hermitian ring laser gyroscopes with enhanced Sagnac sensitivity," *Nature,* vol. 576, no. 7785, pp. 70-74, 2019.

[13] A. Jian *et al.*, "Parity-time symmetry based on resonant optical tunneling effect for biosensing," *Optics Communications,* vol. 475, p. 125815, 2020.

[14] M. De Carlo, F. De Leonardis, and V. M. Passaro, "Design rules of a microscale PT-symmetric optical gyroscope using group IV platform," *Journal of Lightwave Technology,* vol. 36, no. 16, pp. 3261-3268, 2018.

[15] X. Liu, H. Wang, J. Zhang, J. Guo, and X. Wu, "Enhancement of Sensitivity Near Exceptional Point by Constructing Nonreciprocal Fiber Cavity Assisted by Isolator and Erbium-Doped Fiber," *IEEE Sensors Journal,* vol. 21, no. 17, pp. 18823-18828, 2021.

[16] Ş. K. Özdemir, S. Rotter, F. Nori, and L. Yang, "Parity–time symmetry and exceptional points in photonics," *Nature materials,* vol. 18, no. 8, pp. 783-798, 2019.

[17] C. M. Bender, D. C. Brody, and H. F. Jones, "Must a hamiltonian be hermitian?," *American Journal of Physics,* vol. 71, no. 11, pp. 1095-1102, 2003.

[18] C. M. Bender, "Making sense of non-Hermitian Hamiltonians," *Reports on Progress in Physics,* vol. 70, no. 6, p. 947, 2007.

[19] A. K. Jahromi, A. U. Hassan, D. N. Christodoulides, and A. F. Abouraddy, "Statistical parity-time-symmetric lasing in an optical fibre network," *Nature communications,* vol. 8, no. 1, pp. 1-9, 2017.

[20] D. Marcuse, "Coupled mode theory of optical resonant cavities," *IEEE journal of quantum electronics,* vol. 21, no. 11, pp. 1819-1826, 1985.

[21] A. Yariv and P. Yeh, *Photonics: optical electronics in modern communications*. Oxford university press, 2007.

[22] W. Chen, J. Zhang, B. Peng, Ş. K. Özdemir, X. Fan, and L. Yang, "Parity-time-symmetric whispering-gallery mode nanoparticle sensor," *Photonics Research,* vol. 6, no. 5, pp. A23-A30, 2018.